\documentstyle[12pt]{article}

\begin{document}

\begin{flushright}
IMSc/99/04/15 \\
hep-th/9904110
\end{flushright} 

\vspace{2mm}

\vspace{2ex}

\begin{center}
{\large \bf Holographic Principle in the Closed Universe :} \\ 

\vspace{2ex}

{\large \bf a Resolution with Negative Pressure Matter} \\

\vspace{8ex}

{\large  S. Kalyana Rama}

\vspace{3ex}

Institute of Mathematical Sciences, C. I. T. Campus, 

Taramani, CHENNAI 600 113, India. 

\vspace{1ex}

email: krama@imsc.ernet.in \\ 

\end{center}

\vspace{6ex}

\centerline{ABSTRACT}

\begin{quote} 
A closed universe containing pressureless dust, more 
generally perfect fluid matter with pressure-to-density 
ratio $w$ in the range $(\frac{1}{3}, - \frac{1}{3})$, 
violates holographic principle 
applied according to the Fischler-Susskind proposal.  
We show, first for a class of two-fluid solutions and then for 
the general multifluid case, that the closed universe will  
obey the holographic principle if it also contains matter with 
$w < - \frac{1}{3}$, and if the present value of its total 
density is sufficiently close to the critical density. It is 
possible that such matter can be realised by some form of 
`quintessence', much studied recently. 
\end{quote}

\vspace{2ex}


\newpage

\vspace{4ex}

{\bf 1.} 
The holographic principle implies that the degrees of freedom 
in a spatial region can all be encoded on its boundary, with 
a density not exceeding one degree of freedom per planck 
cell \cite{thooft}. Accordingly, the entropy in a spatial 
region does not exceed its boundary area measured in planck 
units. Moreover, the physics of the bulk is describeable by 
the physics on the boundary. This has, indeed, been realised 
recently for some anti de Sitter spaces \cite{sw}. 

Fischler and Susskind (FS) have proposed \cite{fs} how to  
apply the holographic principle in cosmology, and showed  
that our universe, if flat or open, obeys this principle  
as long as its size is non zero - that is, non planckian.  
If closed, however, it violates this principle 
in the future even 
while its size is non zero. In fact, in some cases, 
the violation occurs while the universe is still 
expanding. This may indicate that closed universe is to be  
excluded as inconsistent, or some new behaviour must set 
in to accomodate the holographic principle \cite{fs}. 

The holographic 
principle has since been applied in the context of pre 
big bang scenario \cite{rey1}, singularity problem  
\cite{mag}, and inflation \cite{ks,lowe}, mostly for 
flat universe. Recently, there have been two 
alternative proposals for the implementation of  
the holographic principle: by Easther and Lowe, based on 
second law of thermodynamics \cite{lowe}; and, by 
Bak and Rey, using the `cosmological apparent horizon' 
instead of particle horizon \cite{rey}. 
In both of these implementations,  
the closed universe also obeys the holographic principle 
naturally. Therefore, these proposals are perhaps 
the more natural ones than the FS proposal. 

Nevertheless, it is of interest to study whether or not 
a closed universe is indeed to be excluded as inconsistent  
with the holographic principle, applied according to the FS  
proposal. We study this issue in this paper. 

Throughout in the following we consider a closed universe, 
initially of zero size and expanding. It is assumed to contain  
more than one type of non interacting perfect fluid matter. The 
pressure $p_i$ and the density $\rho_i$ of the $i^{{\rm th}}$ 
type of matter is related by the equation of state  
\begin{equation}\label{eos}
p_i = w_i \rho_i \; \; ,  \; \; \; \; - 1 \le w_i \le 1 \; . 
\end{equation}
The parameter $w$ denotes the nature of the matter: 
$w = 0$ for pressureless dust, $w = \frac{1}{3}$ for radiation, 
etc. Furthermore, we assume that one of the $w$'s, say $w_1$,  
lies in the range 
\begin{equation}\label{w1}
- \frac{1}{3} < w_1 < \frac{1}{3}  
\end{equation}
so that if the corresponding matter were the only one 
present, then the universe violates the holographic 
principle in the future (see \cite{fs}). 

We study the explicit solution for the two-fluid case 
with $w_1 + w_2 = - \frac{2}{3}$. We find that 
the closed universe  
obeys the holographic principle throughout its future if 
and only if the present value of the total density is 
sufficiently close to the critical density. 

Using this solution, we show furthermore that 
{\em the closed universe, containing atleast one matter, 
with its $w$ lying in the range} (\ref{w1}), 
{\em obeys the holographic  
principle, applied according to the FS proposal, if it also  
contains atleast one other matter with its $w$ satisfying  
$w < - \frac{1}{3}$, and if the present value of its total 
density is sufficiently close to the critical density.}  
Thus, the closed universes need not 
be excluded as inconsistent, nor any new behaviour needs 
to set in; the above requirements will suffice. 

If these conditions are also necessary 
then the holographic principle, applied 
according to the FS proposal, can be  said to predict that  
if the total density at present 
of our universe exceeds the critical density, no matter 
by how small an amount, then it is closed and, hence, must  
also contain matter with $w < - \frac{1}{3}$. 

We make a few remarks about matter with negative values 
of $w$. No known physical matter is of this type, except 
cosmological constant ($w = - 1$). However, such 
matter, with $w \le - \frac{1}{\sqrt{3}}$, was found necessary 
in \cite{k97} in avoiding big bang singularity within low 
energy string theory. Furthermore, Dirichlet - $0$ and/or 
$(- 1)$ - branes \cite{pol} were envisaged as possible 
candidates for such matter. Also, the recent discovery, 
through the analyses of distant Supernovae, that the 
universe   is accelerating at present \cite{perl} has sparked  
an enormous  interest in the study of matter with $w < 0$.  
A realisation of such matter is the so called 
`quintessence' - a time varying, spatially  
inhomogeneous component of energy density of the universe 
with negative pressure, much studied  
recently \cite{qneed,q}. 
Some of the references which study various 
candidates for matter with negative pressure and/or 
quintessence are given in \cite{qmodel}. It is possible 
that  matter with $w < - \frac{1}{3}$, required here, 
can be realised by some one of the above candidates.  

In the following, we first outline how the closed universe 
violates the holographic principle \cite{fs} in the future. 
We then present two-fluid solutions, and obtain the conditions  
underwhich the holographic principle is obeyed. We then show 
that these conditions are also valid in general. 

\vspace{4ex}

{\bf 2.} 
The line element for the homogeneous isotropic universe 
is given by 
\[
d s^2 = - d t^2 + R^2 (t) \left( \frac{d r^2}{1 - k r^2} 
+ r^2 d \Omega_2^2 \right) \; , 
\]
where $R$ is the scale factor, $d \Omega_2$ is the standard 
line element on unit
two dimensional sphere, and $ k = 0, - 1$, or $ + 1$ for 
flat, open, or closed universe respectively. The above 
line element can also be written as  
\[
d s^2 = - d t^2 + R^2 (t) (d \chi^2 + r^2 d \Omega_2^2) \; ,  
\]
where $r = \chi, \sinh \chi$, or 
$\sin \chi$ for $k = 0, - 1$, or $ + 1$ respectively. 
The coordinate size of the horizon is given by the parameter 
\[
\chi = \int_0^t \frac{d t}{R} \; . 
\]
The holographic bound is given, upto numerical factors of  
${\cal O}(1)$, by $S \le A$  
where $S$ is the entropy in a given region and $A$ is 
the area of its boundary in Planck units \cite{thooft}. 
Applied to the closed universe 
according to FS proposal, it implies, 
upto numerical factors of ${\cal O}(1)$, that 
\begin{equation}\label{fs}
\frac{S}{A} = \frac{\sigma (2 \chi - \sin 2 \chi)}
{R^2 \sin^2 \chi} \; \le 1  \; , 
\end{equation}
where $\sigma$ is the constant comoving entropy density 
\cite{fs}. 

For a closed universe, initially of zero size and 
expanding, and containing only one type of matter 
with $w > - \frac{1}{3}$, one obtains \cite{kramer}, see 
below also, 
\begin{equation}\label{single}
R^{1 + 3 w} \propto \sin^2 \frac{1 + 3 w}{2} \chi \; .  
\end{equation}
It can be seen that if $w \ge \frac{1}{3}$, then the FS bound 
(\ref{fs}) will be violated only when $R \to 0$, which is the
Planckian regime. However, if 
$- \frac{1}{3} < w < \frac{1}{3}$, then the FS bound (\ref{fs}) 
will
be violated as $\chi \to \pi$. The violation occurs even while
$R$ is non zero.  In fact, when $w \le 0$, the violation occurs
while the universe is still expanding. This may indicate that
such universes are to be excluded as inconsistent, or some new
behaviour must set in to accomodate the holographic principle, 
applied according to the FS proposal \cite{fs}.

\vspace{4ex}

{\bf 3.} 
The above conclusion is valid 
if the universe contains one type of matter only. 
In reality, however, more than one type of matter will be  
present in various, perhaps subdominant, quantities. It is 
then important to study such multifluid solutions before 
excluding closed universes as inconsistent. 

The general multifluid solutions are difficult
to obtain. In a few cases where exist \cite{multi}, they
typically involve elliptic functions and are often not
in a useful form. However, for a class of models,
we now present general solutions, in a form useful 
for our purposes.

Assume that the universe contains different types of 
non interacting perfect fluid matter, with equations 
of state given as in (\ref{eos}). We assume, without loss of 
generality, that $w_i \ne - \frac{1}{3}$ since the effect 
of such matter is same as that of the $k$-term 
in equation (\ref{rdot}) below. Also, define 
\[
\Omega_i \equiv \frac{\rho_{0i}}{\rho_c} 
\; , \; \; \; \; 
\rho_c \equiv \frac{3 H_0^2}{8 \pi G} 
\; , \; \; \; \; 
H_0 \equiv \left( \frac{\dot{R}}{R} \right)_0 
\]
where $\rho_{0i}$ ($\ne 0$), 
is the present value of $\rho_i$, $\rho_c$  
is the critical density, $G$ is the Newton's constant, and 
$H_0$ is the present value of the Hubble parameter. Einstein's  
equations of motion can then be written as 
\begin{eqnarray} 
\dot{R}^2 & = & - k + H_0^2 R_0^2 \sum_i \Omega_i 
\left( \frac{R}{R_0} \right)^{- (1 + 3 w_i)} 
\; \equiv f(R) \label{rdot} \\ 
\rho_i & = & \rho_{0i} 
\left( \frac{R}{R_0} \right)^{- 3 (1 + w_i)} \; , \label{rho}
\end{eqnarray}
where $R_0$ is the present value of $R$, and the upper 
dots denote the time derivatives. The present value of 
$\dot{R}$ gives the relation  
\begin{equation}\label{k} 
k = H_0^2 R_0^2  \; (\sum_i \Omega_i - 1 ) \; . 
\end{equation} 

Throughout in the following, let $w_1$ lie  
in the range given in (\ref{w1}). Thus, 
if the corresponding matter were the only one 
present, then the universe violates the holographic 
principle in the future (see \cite{fs}). 
We now define $y$ and $x$ as follows: 
\begin{eqnarray}
\frac{R}{R_0} & = & \left( H_0^2 R_0^2 \Omega_1 
\right)^{\frac{1}{1 + 3 w_1}} \; \; 
y^{\frac{2 a}{1 + 3 w_1}} \label{y} \\
\frac{d t}{R_0} & = & \frac{2 a}{1 + 3 w_1} \; 
\left( H_0^2 R_0^2 \Omega_1 \right)^{\frac{1}{1 + 3 w_1}} \; \; 
y^{\frac{2 a q}{1 + 3 w_1}} d x \label{t} \; , 
\end{eqnarray}
where $a$ and $q$ are positive constants to be chosen 
suitably, and we set $x = 0$ at $t = \chi = 0$. Clearly, 
the parameter $\chi$ is given by 
\begin{equation}\label{eta} 
\chi = \int_0^t \frac{d t}{R} = 
\frac{2 a}{1 + 3 w_1} \; 
\int_0^x d x \; y^{\frac{2 a (q - 1)}{1 + 3 w_1}} \; . 
\end{equation}
In terms of $y$ and $x$, equation (\ref{rdot}) becomes 
\begin{equation}\label{eqn}
\left( \frac{d y}{d x} \right)^2 = - k y^\alpha 
+ \sum_i c_i y^{\alpha_i} \; \equiv g(y) \; , 
\end{equation} 
where the exponents $\alpha$ and $\alpha_i$, and the constants  
$c_i$, are given by 
\begin{eqnarray}
\alpha & = & 2 + \frac{4 a (q - 1)}{1 + 3 w_1} 
\label{alpha} \\ 
\alpha_i & = & \alpha - \frac{2 a (1 + 3 w_i)}{1 + 3 w_1} 
\label{alphai} \\ 
c_i & = & H_0^2 R_0^2 \Omega_i \; \left( H_0^2 R_0^2 
\Omega_1 \right)^{- \frac{1 + 3 w_i}{1 + 3 w_1}} \; . 
\label{ci} 
\end{eqnarray}
Note that $c_1 = 1$. The function $f(R)$ in (\ref{rdot}), 
expressed in terms of $y$, becomes $f(R) = y^{- \alpha} g(y)$.  

From now on, we set $k = + 1$. If $q = a = 1$ then 
$(\alpha, \alpha_1) = (2, 0)$, and the solution 
(\ref{single}) for the single fluid case \cite{kramer} 
follows trivially. Consider two-fluid cases: 
$i = 1, 2$. The boundary conditions, corresponding to  
an universe, initially of zero size and expanding, are  
$R = y = 0$ and $\dot{R} > 0$ at $t = x = 0$. 

\vspace{2ex}

{\bf A:} 
Let $q = a = 1$, and $(1 + 3 w_1) = 2 (1 + 3 w_2)$. 
To be definite, let $(w_1, w_2) = (\frac{1}{3}, 0)$ and 
$(\Omega_1, \Omega_2) = (\Omega_r, \Omega_d)$ denoting   
radiation and pressureless dust respectively. 
Then $(\alpha, \alpha_1, \alpha_2) = (2, 0, 1)$, 
and equation (\ref{eqn}) becomes 
\[
\left( \frac{d y}{d x} \right)^2 = 1 + c y - y^2 
\; \; , \; \; \; \; 
c = \frac{H_0 R_0 \Omega_d}{\sqrt{\Omega_r}} \; . 
\]
The solution for $R$ is given,  
after a straightforward algebra, by  
\begin{equation}\label{rrad}
R = A R_0 \; (\sin (\chi - \alpha) + \sin \alpha) 
\; \; , \; \; \; \; 
t = \int_0^\chi d \chi \; R \; ,  
\end{equation}
where the constants $A$ and $\alpha$ are given by 
\[
4 A^2 = H_0^2 R_0^2 \; 
(4 \Omega_r + H_0^2 R_0^2 \Omega_d^2) 
\; \; , \; \; \; \; \; 
\tan \alpha = \frac{c}{2} \; . 
\]
It follows from the above expressions that 
\[
R \vert_{\chi = \pi} = 2 A R_0 \; \sin \alpha = 
H_0^2 R_0^3 \Omega_d \; . 
\]
Hence, in a closed universe with both radiation  
and dust present, the FS bound (\ref{fs}) will be violated 
even while $R$ ($= R \vert_{\chi = \pi}$) is non 
zero. This is true irrespective of the amount of radiation 
present, however large it may be. Only when $\Omega_d = 0$ 
exactly, will the FS bound (\ref{fs}) 
be obeyed all the way until $R = 0$, 
{\em i.e.} until the universe recollapses to zero size.

\vspace{2ex}

{\bf B:} 
Let $2 a = 1$, and 
$2 (q - 1) = - (1 + 3 w_1) = (1 + 3 w_2)$. 
(For example, $(w_1, w_2) = (0, - \frac{2}{3})$.) 
Then $(\alpha, \alpha_1, \alpha_2) = (1, 0, 2)$, 
and equation (\ref{eqn}) becomes 
\[
\left( \frac{d y}{d x} \right)^2 = 1 - y + c y^2 
\; \; , \; \; \; \; 
c = H_0^4 R_0^4 \; \Omega_1 \Omega_2 \; . 
\]
Using equation (\ref{k}), the constant $c$ can be 
written as 
\begin{equation}\label{c}
c = \frac{\Omega_1 \Omega_2}{(\Omega_1 + \Omega_2 - 1)^2} \; ,  
\end{equation}
where $\Omega_1 + \Omega_2 > 1$ since $k = + 1$. 
The parameter $\chi$ and the time $t$ are given by  
\begin{eqnarray}
\chi & = & \frac{1}{1 + 3 w_1} \; 
\int_0^x \frac{d x}{\sqrt{y}} \label{chi2} \\
t & = & \frac{R_0}{1 + 3 w_1} \;  \left( H_0^2 R_0^2 
\Omega_1 \right)^{\frac{1}{1 + 3 w_1}} \; 
\int_0^x \frac{d x}{\sqrt{y}} \; 
y^{\frac{1}{1 + 3 w_1}} \; .  \label{etat2}
\end{eqnarray}
The details of the solution depend on whether or not 
$f_{{\rm Min}} = {\rm Min} (\frac{1}{y} - 1 + c y) = 
(2 \sqrt{c} - 1)$
is negative or positive. Consider now each of these cases.     

\vspace{2ex}

\centerline{ \underline{ $f_{{\rm Min}} < 0$} 
$\; \; \longleftrightarrow \; \; $ 
\underline{$2 \sqrt{c} < 1$ }  }

\vspace{2ex}

The solution for $y$ is given, 
after a straightforward algebra, by  
\begin{equation}\label{yc<}
y \sqrt{c} \sinh \alpha = \cosh \alpha 
- \cosh (x \sqrt{c} - \alpha) \; \; , \; \; \; \; 
\tanh \alpha \equiv 2 \sqrt{c} \; .
\end{equation}
Thus, $y$ starts from zero at $x = 0$, expands to a maximum,  
given by $y_{{\rm max}} \sqrt{c} = \tanh \frac{\alpha}{2}$, 
at $x \sqrt{c} = \alpha$, and then  recollapses to zero at 
$x \sqrt{c} = 2 \alpha$. It can be seen from equation 
(\ref{etat2}) that the recollapse occurs in a finite time.   

The parameter $\chi$ is given, after a straightforward algebra  
using equation (\ref{chi2}) and the formula (2.464(32)) of 
\cite{grad}, by 
\begin{equation}\label{etac<}
\frac{1 + 3 w_1}{2} \; \chi = 
\sqrt{\frac{2 \cosh \alpha}{1 + \cosh \alpha}} \; \; 
F(\phi, \beta) \; , 
\end{equation}
where 
\begin{equation}\label{fphibeta}
F(\phi, \beta) = \int_0^\phi 
\frac{d \theta}{\sqrt{1 - \beta^2 \sin^2 \theta}}
\end{equation}
is the elliptic integral of the first kind. 
The parameters $\phi$ and $\beta$ are given by 
\[
\sin^2 \phi = \frac{y}{y_{{\rm max}}}
\; \; , \; \; \; \; 
\beta = \tanh \frac{\alpha}{2} \; < 1 \; . 
\]
Thus, as $y$ expands from zero to $y_{{\rm max}}$ and then
recollapses to zero, $\phi$ increases monotonically 
from $0$ to $\frac{\pi}{2}$ to $\pi$. 
Also, the parameter $\chi$ increases monotonically 
with $\phi$ and, since $\beta^2 < 1$, remains finite always.   

It is now important to find out whether or not 
the value of $\chi$ at the time of recollapse, 
$\chi_* \equiv \chi \vert_{\phi = \pi} > \pi$. Clearly, 
if $\chi_* > \pi$ then the FS bound (\ref{fs}) will be violated 
even while $R$ is non zero. If $\chi_* \le \pi$ then the FS  
bound (\ref{fs}) will be obeyed 
all the way until $R = 0$, {\em i.e.} until  
the universe recollapses to zero size. Towards this end, 
note from equations (\ref{etac<}) and (\ref{fphibeta}) that 
\[
\frac{1 + 3 w_1}{2} \; \chi > \phi 
\; \; \; \; {\rm and,} \; {\rm hence,} \; \; \; \; 
\chi_* > \frac{2 \pi}{1 + 3 w_1} \; . 
\]
Thus, for $w_1 < \frac{1}{3}$, 
$\chi_* > \pi$ and, therefore, the FS bound (\ref{fs}) will be 
violated even while $R$ is non zero. In fact, for dust 
($w_1 = 0$), the violation occurs while the universe 
is still exapnding; that is, $\chi = \pi$ even while 
$\phi < \frac{\pi}{2}$.

\newpage

\vspace{2ex}

\centerline{ \underline{ $f_{{\rm Min}} > 0$} 
$\; \; \longleftrightarrow \; \; $ 
\underline{$2 \sqrt{c} > 1$ }  }

\vspace{2ex}

The solution for $y$ is given, 
after a straightforward algebra, by  
\begin{equation}\label{yc>}
y \sqrt{c} \cosh \alpha = \sinh (x \sqrt{c} - \alpha) 
+ \sinh \alpha  \; \; , \; \; \; \; 
\tanh \alpha \equiv \frac{1}{2 \sqrt{c}} \; .
\end{equation}
Thus, $y$ starts from zero at $x = 0$, expands to infinity 
as $x \to \infty$. It can be seen from equation 
(\ref{etat2}) that, for $w_1 < \frac{1}{3}$, the required  
time $t$ also $\to \infty$. 

The parameter $\chi$ is given, after a straightforward algebra  
using equation (\ref{chi2}) and the formula (2.464(16)) of 
\cite{grad}, by 
\begin{equation}\label{etac>}
(1 + 3 w_1) \chi = c^{- \frac{1}{4}} \; F(\phi, \beta) \; , 
\end{equation}
with $F(\phi, \beta)$ as given in (\ref{fphibeta}). 
The parameters $\phi$ and $\beta$ are now given by 
\[
\cos \phi = \frac{1 - y \sqrt{c}}{1 + y \sqrt{c}} 
\; \; , \; \; \; \; 
\beta^2 = \frac{1 + \tanh \alpha}{2} \; < 1 \; . 
\]
Thus, as $y$ expands from $0$ to infinity, $\phi$ increases 
monotonically from $0$ to $\pi$. Also, $\chi$ increases 
monotonically with $\phi$ and, since $\beta^2 < 1$, remains  
finite always. 

For the same reasons as given before, it is now important to 
find out whether or not the value of $\chi$ as $y \to \infty$, 
$\chi_* \equiv \chi \vert_{\phi = \pi} > \pi$. Clearly if 
$\chi_* > \pi$ then the FS bound (\ref{fs}) will be violated 
even while $R$ is non zero and, in fact, increasing. If  
$\chi_* < \pi$ then the FS bound (\ref{fs}) 
will be obeyed for all times 
$t$. Towards this end, note from equation (\ref{etac>}) that   
\begin{equation}\label{eta*}
\chi_* = \frac{2 c^{- \frac{1}{4}}}{1 + 3 w_1} \; K(\beta)  
\end{equation}
where we have used $F(\pi, \beta) = 
2 F(\frac{\pi}{2}, \beta) \equiv 2 K (\beta)$. Here, 
$K(\beta)$ is the complete 
elliptic integral, and is finite since $\beta^2 < 1$. Thus,  
it follows from equation (\ref{eta*}), the $c$-dependence of 
$\beta$, and the properties of $K(\beta)$, that 
$\chi_* < \pi$ implies that $c > c_*$, where $c_*$ is the 
solution of equation (\ref{eta*}) when $\chi_* = \pi$. 
This, in turn, implies that 
\[
f_{{\rm Min}} > f_* \; \equiv 2 \sqrt{c_*} - 1 
\]
and also, from equation (\ref{c}), that 
$\Omega_2 < \Omega_{2*}$ for a given value of $\Omega_1$.  

It also follows from equation (\ref{eta*}), the $c$-dependence 
of $\beta$, and the properties of $K(\beta)$, that $c_*$ 
increases and, hence, $\Omega_{2*}$ decreases, as $w_1$ 
decreases.  However, an explicit expression for $c_*$ is not 
available, although one can approximately determine $c_*$ using 
the tabulated values of $K (\beta)$. Then, $\Omega_{2*}$ can be 
determined for a given $\Omega_1$.

For example, if $w_1 = 0$ then $c_* \approx 2.684$, and  
\[
(\Omega_1, \Omega_{2*}) \approx (0.1, 1.103), \; \; 
(0.3, 1.041), \; \; (0.5, 0.912), \; \; \cdots 
\]
If $w_1 = \frac{1}{3}$, the highest value possible in 
the present solution, then $c_* \approx 0.416$, and 
\[
(\Omega_1, \Omega_{2*}) \approx (0.1, 1.501), \; \; 
(0.3, 1.857), \; \; (0.5, 2.082), \; \; \cdots 
\]
In all these cases, we have 
$\Omega_1 + \Omega_{2*} = 1 + {\cal O}(1)$. Thus, we see 
that $\chi_* < \pi$ and, hence, the FS bound (\ref{fs}) 
will be obeyed if matter, with $w < - \frac{1}{3}$,  
is present and if 
\[
0 < \Omega_1 + \Omega_2 - 1 \stackrel{<}{_\sim} 
{\cal O}(1) \; .
\]

\vspace{4ex}

{\bf 4.}
The above result is obtained for the two-fluid solutions 
where $(1 + 3 w_1) = - (1 + 3 w_2)$. However, this 
result is valid for general multifluid solutions also.  
Namely, the FS bound (\ref{fs}) will be obeyed if 

\noindent
{\bf (1)} atleast one matter, with $w < - \frac{1}{3}$,  
is present,  and 

\noindent
{\bf (2)} the present value of the total density is 
sufficiently close to the critical density, {\em i.e.} 
$(\sum_i \Omega_i - 1)$, which must be positive since 
$k = + 1$, is sufficiently small. 

This can be proved as follows. Let the multifluid 
system contain atleast two types of matter, one 
with its $w \equiv w_1 > - \frac{1}{3}$, and another 
with its $w \equiv w_2 < - \frac{1}{3}$. It may now contain 
other types of matter also, with no further restrictions on 
$\{ w_i \}$, $i = 1, 2, \cdots \;$. Then, the function 
\begin{equation}\label{h}
h(R) \equiv \sum_i \Omega_i \left( \frac{R}{R_0} 
\right)^{- (1 + 3 w_i)} 
\end{equation} 
has its non zero, and the only, minimum 
at a finite value of $R$. That is,  
\[
h(R) \ge h(R_m) > 0 
\; \; , \; \; \; \; 
0 < R_m < \infty \; . 
\]

We now consider an auxiliary two-fluid system, with the 
corresponding function given by 
\begin{equation}\label{haux}
\tilde{h}(R) \equiv \sum_{j = 1, 2}
\tilde{\Omega}_j \left( \frac{R}{R_0} 
\right)^{- (1 + 3 \tilde{w}_j)} \; ,  
\end{equation} 
where the tildes refer to the auxiliary system. By a simple, 
but slightly involved, analysis  
it can be shown that the 
parameters $\tilde{w}_j$ and $\tilde{\Omega}_j$, 
$j = 1, 2$, can be chosen 
\footnote{For example, choose $\tilde{w}_1$ such that  
${\rm Min} \{ \vert 1 + 3 w_i \vert \} > 
1 + 3 \tilde{w}_1 > 0$,  
and $\tilde{\Omega}_j$'s such that $\tilde{h}(R)$ also has 
its non zero, and the only, minimum at $R = R_m$ and 
$h(R_m) > \tilde{h}(R_m) > 0$, 
where the inequalities are obeyed by sufficient margins.} 
such that 
$(1 + 3 \tilde{w}_1) = - (1 + 3 \tilde{w}_2) > 0 $ and 
\[
h(R) > \tilde{h}(R) 
\; \; , \; \; \; \; 
0 \le R \le \infty \; . 
\]
The parameter $H_0 R_0$, taken to be the same for  
both systems, is given by (see equation (\ref{k})) 
\[
H_0^2 R_0^2  \; (\sum_i \Omega_i - 1) = 1 \; . 
\]
The solution for the auxiliary system is nothing but the one 
given in the previous section, where now the parameter 
$\tilde{c} = H_0^4 R_0^4 \tilde{\Omega}_1 \tilde{\Omega}_2$. 

Note that 
\[
f(R) = - 1 + H_0^2 R_0^2 \; h(R) 
\; \; \; \; {\rm and} \; \; \; \; 
\chi = \int_0^t \; \frac{d t}{R} = 
\int_0^R \frac{d R}{R \sqrt{f}} \; , 
\]
and similarly for $\tilde{f}(R)$ and $\tilde{\chi}$. 
Since $h(R) > \tilde{h}(R)$ and $H_0 R_0$ is same 
for both the systems, it follows that 
\[
f(R) > \tilde{f}(R) 
\; \; \; \; {\rm and,} \; \; {\rm hence,} \; \; \; \;  
\chi < \tilde{\chi} \; . 
\]
However, it is shown in the previous section that 
$\tilde{\chi} < \pi$ if $\tilde{c}$ is sufficiently  
large. Therefore, it now follows that the parameter 
\[
\chi < \pi 
\]
and, hence, the FS bound (\ref{fs}) is obeyed in the 
future, if $(\sum_i \Omega_i - 1)$, which must be positive, 
is sufficiently small; that is, if the present value of 
the total density is sufficiently close to the critical 
density. 

We have thus shown that 
{\em the closed universe, containing atleast one matter, 
with its $w$ lying in the range} (\ref{w1}), 
{\em obeys the holographic  
principle, applied according to the FS proposal, if it also  
contains atleast one other matter with its $w$ satisfying  
$w < - \frac{1}{3}$, and if the present value of its total 
density is sufficiently close to the critical density.}  
Thus, the closed universes need not 
be excluded as inconsistent, nor any new behaviour needs 
to set in; the above requirements will suffice. 

If these conditions are also necessary, then they can be 
taken as predictions of holographic principle, applied to 
the closed universe according to the FS proposal. 
Thus, if the total density at present 
of our universe, which certainly 
contains pressureless dust, exceeds 
the critical density, no matter by how small an amount, 
then it is closed, $k = + 1$. The holographic principle, 
applied according to tha FS proposal, would then require  
that our universe must also contain matter with 
$w < - \frac{1}{3}$. It is 
possible that such matter can be realised by some form of 
quintessence, much studied recently.


\end{document}